\documentclass[aps,showpacs,a4paper,floatfix,twocolumn,prc,amsmath,amssymb]{revtex4-1}
\usepackage{graphicx}
\usepackage{epsfig}
\usepackage{ulem}
\usepackage{morefloats}
\usepackage{rotating}
\usepackage{relsize}
\usepackage{color}      

\usepackage{url}

\usepackage{amsmath}

\begin{document}

\title{Is the magnetic field making the inner crust of neutron stars more heterogeneous?}

\author{Jianjun Fang$^1$,Helena Pais$^1$} 
\author{Sidney Avancini$^{1,2}$}
\author{Constan\c ca Provid{\^e}ncia$^1$}

\affiliation{$^1$CFisUC, Department of Physics, University of Coimbra, 3004-516 Coimbra, Portugal.\\
$^2$Departamento de F\'{\i}sica, Universidade Federal de Santa Catarina, Florian\'opolis, SC, CP. 476, CEP 88.040-900, Brazil
} 

\begin{abstract}

We study the effect of strong magnetic fields, of the order of
$10^{15}-10^{17}$ G, on the extension of the crust of
magnetized neutron stars. The  dynamical instability region
of neutron-proton-electron ($npe$) 
matter at subsaturation
densities and the  mode with the largest growth rate are determined within a relativistic mean field model. It is
shown that the effect of a strong magnetic field on the instability region is very sensitive to the
density dependence of the symmetry energy, and that it originates an 
increase of the extension of the crust and of the charge content of clusters.
\end{abstract}

\maketitle

Soft $\gamma$-ray repeaters and some anomalous X-ray pulsars are
strongly magnetized neutron stars  known as magnetars
\cite{duncan,usov,pacz}. These stars  have strong surface  magnetic
fields of the order of $10^{14}-10^{15}$ G \cite{sgr}, and slow
rotation with a period of $\sim 1-12$ s. 
Recently,  the time evolution of the magnetic
field of  isolated
X-ray pulsars has been studied by Pons et al. \cite{pons13}. 
 The authors
have shown that a fast decay of the magnetic field  could explain the  non observation of stars with periods above
12 s. The decay of the magnetic field was obtained including
 a high electrical resistivity in the inner crust,  attributed to the
 possible
 existence of an amorphous and heterogeneous layer at the bottom of
the inner crust. The lack of isolated X-ray pulsars with a
period {higher than} $12$s, could, therefore, be a direct indication of the
existence of an amorphous inner crust, possibly in the form of 
pasta phases \cite{pons13}.

At low nuclear matter
densities, a competition between the long-range Coulomb repulsion and
short-range nuclear attraction will lead to the formation of 
clusterized matter,  known as nuclear pasta \cite{pasta},  near the crust-core transition.  
These geometrical configurations are observed not only in
nuclear matter, but in a variety of amorphous solids, crystals, and
magnetic and biological materials \cite{materials}.
One of the
main interests on the existence of these exotic structures in the crust of neutron stars is the effect that they might have on the neutrino transport, and subsequent cooling of the neutron star \cite{neutrinos}.

Molecular dynamics simulations of
the nuclear pasta have shown that topological defects in the pasta could increase
electron scattering and reduce  the  electrical and the thermal
conductivities \cite{horowitz15}. Electron conductivity in magnetized
neutron star matter was also studied in \cite{yakovlev15}, and it was shown
that the electron transport is strongly anisotropic, due to the
presence of strong magnetic fields. The complexity introduced
by the magnetic field suggests that both suppression and
enhancement of the electron conduction in the presence of the pasta
phases are possible, and further calculations are required.

Stellar matter contains, besides neutrons and protons, also electrons,
that neutralize the proton charge. The
transition clusterized-homogeneous matter has been estimated using
different methods. In particular, in \cite{avancini08}, it has been
shown that a Thomas-Fermi (TF) description of the pasta phase predicts the
same crust-core transition density as a
dynamical spinodal calculation, which allows  independent electron
and proton density fluctuations. 
The same conclusion was drawn in \cite{Avancini-10}, where it
  was shown that the dynamical spinodal calculation gives a lower
  limit on the crust-core density, and that the larger the isospin
  asymmetry, the closer this value is to the TF result. For $\beta-$equilibrium
  matter, both results are practically coincident.

The  importance of the thermodynamical and dynamical spinodals on the
  determination of the behavior of a system that enters an
  instability region was also pointed out to be connected with nuclear multifragmentation, in particular the time evolution of a compound
  nucleus during a heavy ion collision. These instabilities are
  associated to the density region where the curvature of the free
  energy  is negative (\cite{Mueller2005},\cite{Chomaz2004}).

Very strong magnetic fields will influence the proton  charge
fluctuations, and correspondingly the transport properties \cite{yakovlev15}. This
effect was not considered in the above studies.
We will study in the present Letter the effect of a strong magnetic
field on collective modes of stellar matter and dynamical instabilities. 
This is the first time that such a study is being considered,
  as far as we know. In a previous work, some of the authors
have   studied the effect of strong magnetic fields in the nuclear pasta
  phase, using relativistic mean field (RMF) models within a TF
  calculation. However, only magnetic fields above $10^{17}$G  and
  large proton fractions were considered \cite{Lima13} and the anomalous
magnetic moments (AMM) of protons and neutrons were neglected. Moreover, some results showed abrupt behaviors, which were
not totally understood.

Here, we will restrict ourselves to the longitudinal modes arising from small
oscillations around a stationary state in  asymmetric
nuclear matter at subsaturation densities. This investigation will be performed in the framework
of a relativistic mean field hadronic model within the Vlasov
formalism \cite{nielsen91,umodes06,umodes06a}. We will study the effect of the magnetic field on the
spinodal section, the clusterized-homogeneous matter transition, and
the average size of clusters in the nonhomogeneous phase.

Stellar matter is described within the nuclear relativistic mean-field
formalism under the effect of strong magnetic fields
\cite{broderick,aziz08}, including the effect of the AMM. Nucleons with mass $M$ interact with and through an
isoscalar-scalar field $\phi$ with mass $m_s$, an isoscalar-vector
field $V^{\mu}$ with mass $m_v$, and an isovector-vector field
$\mathbf b^{\mu}$ with mass $m_\rho$. Besides nucleons, electrons will
also be included in the Lagrangian density.
Protons and electrons interact through the electromagnetic field
$A^{\mu}$, which includes the static component
$
 A_{\mbox{\tiny stat}}^{\mu}=(0,0,Bx,0),
$
so that $\bf B$=$B$ $\hat{z}$ and $\nabla \cdot {\bf A}$=0. The
static electromagnetic field is assumed to be externally generated,
and only frozen-field configurations will be considered for this component.

The Lagrangian density (taking $c=\hbar=$1) can be written as
$
{\cal L}=\sum_{i=p,n} {\cal L}_i + {\cal L}_e + \cal L_\sigma + {\cal L}_\omega + {\cal L}_\rho + {\cal L}_{\omega\rho} + {\cal L}_{A},
$
with
\begin{eqnarray}
{\cal L}_i&=&\bar \psi_i\left[\gamma_\mu i D^\mu-M^*-\frac{1}{2}\mu_N\kappa_i\sigma_{\mu \nu} F^{\mu \nu}\right]\psi_i,\\
{\cal L}_e&=&\bar \psi_e\left[\gamma_\mu\left(i\partial^\mu + e A^\mu\right)-m_e\right]\psi_e,\nonumber
\end{eqnarray}
where 
$
iD^\mu=i \partial^\mu-g_v V^\mu-
\frac{g_\rho}{2}\boldsymbol\tau \cdot \mathbf{b}^\mu - e A^\mu
\frac{1+\tau_3}{2}$, $M^*=M-g_s\phi,$ 
$e=\sqrt{4\pi/137}$ is  the electromagnetic coupling constant and
$\tau_{3}=\pm 1$ is the
isospin projection for  protons and neutrons,
respectively. The nucleon AMM is
introduced via the coupling of the baryons to the electromagnetic
field tensor with $\sigma_{\mu\nu}=\frac{i}{2}\left[\gamma_{\mu},
  \gamma_{\nu}\right] $ and strength $\kappa_{i}$, with
$\kappa_{n}=-1.91315$ for the neutron, and $\kappa_{p}=1.79285$ for the
proton, and $\mu_N$ is the nuclear magneton. As discussed in
\cite{duncan00}, the contribution of the AMM of
electrons is negligible and will not be considered. For the nuclear
matter parameters, we will consider the NL3
\cite{nl3} and NL3$\omega\rho$ \cite{hor01} parametrizations, with the
symmetry energy slope $L=118$ and 55 MeV, respectively. NL3$\omega\rho$, besides fulfilling several experimental constraints  imposed in Ref. \cite{Dutra}, satisfies within a 10\% deviation  the  constraints imposed by microscopic neutron matter calculations \cite{mic}, and  describes a 2 $M_\odot$ star \cite{Fortin16}.

We will determine the dynamical spinodal within the Vlasov formalism discussed in  \cite{umodes06,nielsen91}. We denote by
$f({\bf r},{\bf p},t)=\mbox{diag}(f_{p},\,f_{n},\,f_{e})$
the  distribution function for $npe$ matter at position $\mathbf r$,
instant $t$ and momentum $\mathbf{p}$,  and by
$
h=\mbox{diag}\left(h_{p},h_{n},h_{e}\right)
$
the corresponding one-body hamiltonian, where
$h_i=\sqrt{\left(\bar{\boldsymbol{p}}_z^i\right)^2+\bar m_i^2} +
{\cal V}^i_{0},$ $i=n,p,e$,
with $\bar{\boldsymbol{p}}^i=\boldsymbol{p-\mathcal{V}}^i$, $
\bar m_p=\sqrt{M^{* 2}_p+2\nu eB}-s\mu_N\kappa_p B$, $\bar m_n=\sqrt{M^{*
    2}_n+\left( \bar{\boldsymbol{p}}_\perp^n\right)^2}-s\mu_N\kappa_nB$, 
$\bar m_e=\sqrt{m^{* 2}_e+2\nu eB}$, ${{{\cal V}^n_\mu}}= g_v {V}_\mu -\frac{g_\rho}{2}\,{ b}_\mu$, ${\cal
  V}^p_{\mu}= g_v V_\mu  + \frac{g_\rho}{2}\, b_\mu+ e\, A_\mu$, 
 ${\cal  V}^e_{\mu}= - e\, A_\mu$, and
$\nu=n+\frac{1}{2}-{\rm sgn}(q)\frac{s}{2}=0, 1, 2, \ldots$ enumerates
the Landau levels of the fermions with electric charge $q$, the
quantum number $s$ is $+1$  (-1) for spin parallel (anti-parallel) to
the magnetic field direction, taken in the
$z$-direction. We define the vectors
(${\boldsymbol{p}},\,{\boldsymbol{V}},...$) along directions parallel ($\boldsymbol{p}_z,\,\boldsymbol{V}_z,...$) and
perpendicular (${\boldsymbol{p}_{\perp}},\,
{\boldsymbol{V}_{\perp}},...$) to the magnetic field.
The time evolution of the distribution function is described by the
Vlasov equation
\begin{equation}
\frac{\partial f_i}{\partial t} +\{f_i,h_i\}=0,\;i=p,\,n,\,e,
\label{vla}
\end{equation}
where $\{,\}$ denote the Poisson brackets.
From the Euler-Lagrange formalism we derive the equations describing the time
evolution of the fields $\phi$,  $V^\mu$, $A^\mu$ and the third component of
the $\rho$-field $b_3^\mu=(b_0,\mathbf{b})$.

At zero temperature, the state which minimizes the energy of asymmetric nuclear matter
is characterized by the Fermi momenta $P_{F}^i$, $i=p,n,e$ and is described by the distribution
function 
$f_0({\bf r},{\bf p}) = \mbox{diag}[\Theta(P_F^{p2}-p^2),\Theta(P_F^{n2}-p^2),\Theta(P_F^{e2}-p^2)],$
where
$P^p_F, \, P^n_F, P^e_F$,
are the Fermi momenta of protons, neutrons and electrons,
and by the constant mesonic field equations.

Collective modes correspond to small oscillations
around the equilibrium state. These small deviations are described by the
linearized equations of motion and the collective modes are the
solutions of those equations. Let the deviations from equilibrium be
described by
$
f\,=\, f_{0} + \delta f\;
$,
$\phi\,=\,\phi_0 + \delta\phi\;$,
$V_0\,=\, V_0^{(0)} + \delta V_0\;$,
$V_i\,=\,\delta V_i\;$, 
$ b_0\,=\, b_0^{(0)} + \delta b_0\;$, 
$b_i\,=\,\delta b_i\;$,
$A_0\,=\, \delta A_0\;$,
$A_i\,=A_{i,\mbox{\tiny stat}}+\,\delta A_i.$
The linearized Vlasov equations for $\delta f_{i}$
are equivalent to the following equations \cite{nielsen91}:
\begin{equation}
\frac{\partial S_i}{\partial t} + \{S_i,h_{0i}\} = \delta h_i, \; i=p,\,n,\,e \, ,
\label{eqevo}
\end{equation}
where $S_i$ are the components of a  generating function 
$S({\bf r},{\bf p}) = \mbox{diag}(S_p, S_n, S_e),$ such that $\delta f_i \,=\, \{S_i,f_{0i}\}$.

In the present work, we consider the longitudinal modes, with momentum
$\boldsymbol{k}$ in the direction of the magnetic field and frequency $\omega$,
described by the ansatz
\begin{equation}
\left(\begin{array}{c}
S_{j}({\bf r},{\bf p},t)  \\
\delta\phi  \\
\delta{\cal B}^\mu 
\end{array}  \right) =
\left(\begin{array}{c}
{\cal S}_{\omega}^j (p,{\rm cos}\theta) \\
\delta\phi_\omega \\
\delta{\cal B}_\omega^\mu
\end{array} \right) {\rm e}^{i(\omega t - {\bf k}_z\cdot
{\bf r})} \;  ,
\label{ans}
\end{equation}
where $j=p,\, n\,, e$, ${\cal B}=V,\, b, A$ represent the vector  fields, and $\theta$ is the angle between ${\boldsymbol p}$ and ${\boldsymbol{k}}_z$.
 For these modes, we have $\delta V_\omega^z = \delta V_\omega$, $\delta b_\omega^z = \delta
b_\omega$ and $\delta A_\omega^z = \delta A_\omega$.
A set of five independent
equations of motion are
obtained, in terms of the amplitudes $A^i_{\omega,j}$ for the proton
and neutron scalar density fluctuations, and the proton, neutron and
electron vector density fluctuations.
The eigenmodes $\omega$ of the system are the solutions of the
dispersion relation obtained, equating to zero the  determinant of the matrix of
the coefficients of the five equations of motion.

\begin{figure*}[t!]
\includegraphics[width=0.80\textwidth]{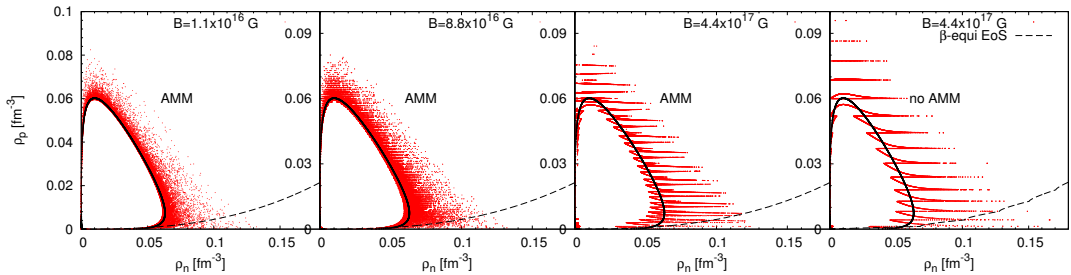} \\
\caption{Dynamical spinodals for the NL3 parametrization, for different magnetic field intensities  and momentum transfer $k=50$ MeV.}
\label{spin}
\end{figure*}

The region in ($\rho_p,\rho_n$) space for a given wave vector $\boldsymbol
k$, satisfying $\omega=0$, defines the dynamical spinodal
surface. At low densities, the system presents unstable modes, characterized by
an imaginary frequency.
Inside the unstable region, 
the mode with
the largest growth rate  $\Gamma$, such  that  $\omega=i\Gamma$,
  is the one that  drives the system to the formation of instabilities. Its
half-wavelength gives a good estimation of the most probable size of the clusters
(liquid) formed in the mixed (liquid-gas) phase \cite{umodes06a}. 
In the following,  possible effects of
strong magnetic fields on the structure of 
the inner  crust of magnetars are discussed from the analysis of the dynamical
spinodal surface and  the unstable modes with the largest growth rate. This approach takes into
account finite size effects, such as the surface tension and Coulomb effects.

{The most intense fields detected on the surface of
a magnetar are not larger than $2\times 10^{15}$ G, the smallest field
we will consider. We may, however, expect stronger fields
in the interior. In particular,  toroidal fields more intense than 10$^{17}$G have
been obtained in stable configurations \cite{kiuchi08,rezzolla12}.
}

In Fig. \ref{spin}, we show the spinodal sections in the
($\rho_p$,$\rho_n$)  space obtained with the NL3 parametrization for the  magnetic fields 
$1.1 \times10^{16}$~G,  $8.8 \times
10^{16}$~G,  and  $4.4 \times10^{17}$ G,  calculated for a wave number
$k=50$ MeV, which gives a spinodal section close to the envelope of
all spinodal sections. 
The calculations were carried
out including AMM, except for the largest field, for which we also
  show the no AMM spinodal. 
{ Due to a numerical limitation, the  spinodal sections are
made of points, which, however, define close regions. Each point is a
solution of dispersion relation obtained for a fixed proton fraction
which varies between 0 an 1.}
 The thick black line
represents the spinodal section for a zero magnetic field. The
dashed line is the EoS of $\beta$-equilibrium matter, and allows the
identification of the crust-core transition.

\begin{figure}[t!]
\begin{tabular}{cc}
\includegraphics[height=7.6cm]{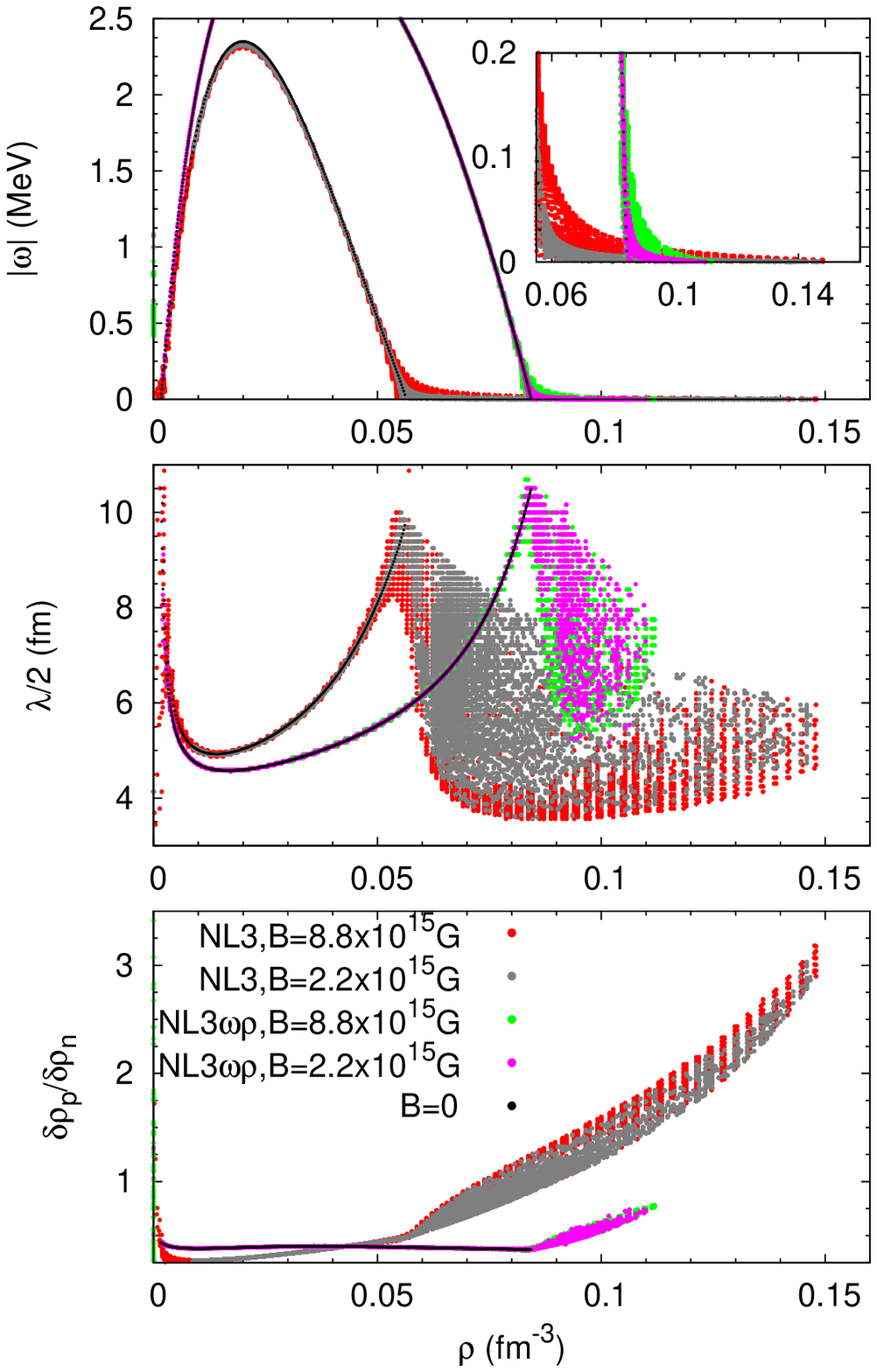} & %
\hspace{-0.cm}
\includegraphics[height=7.6cm]{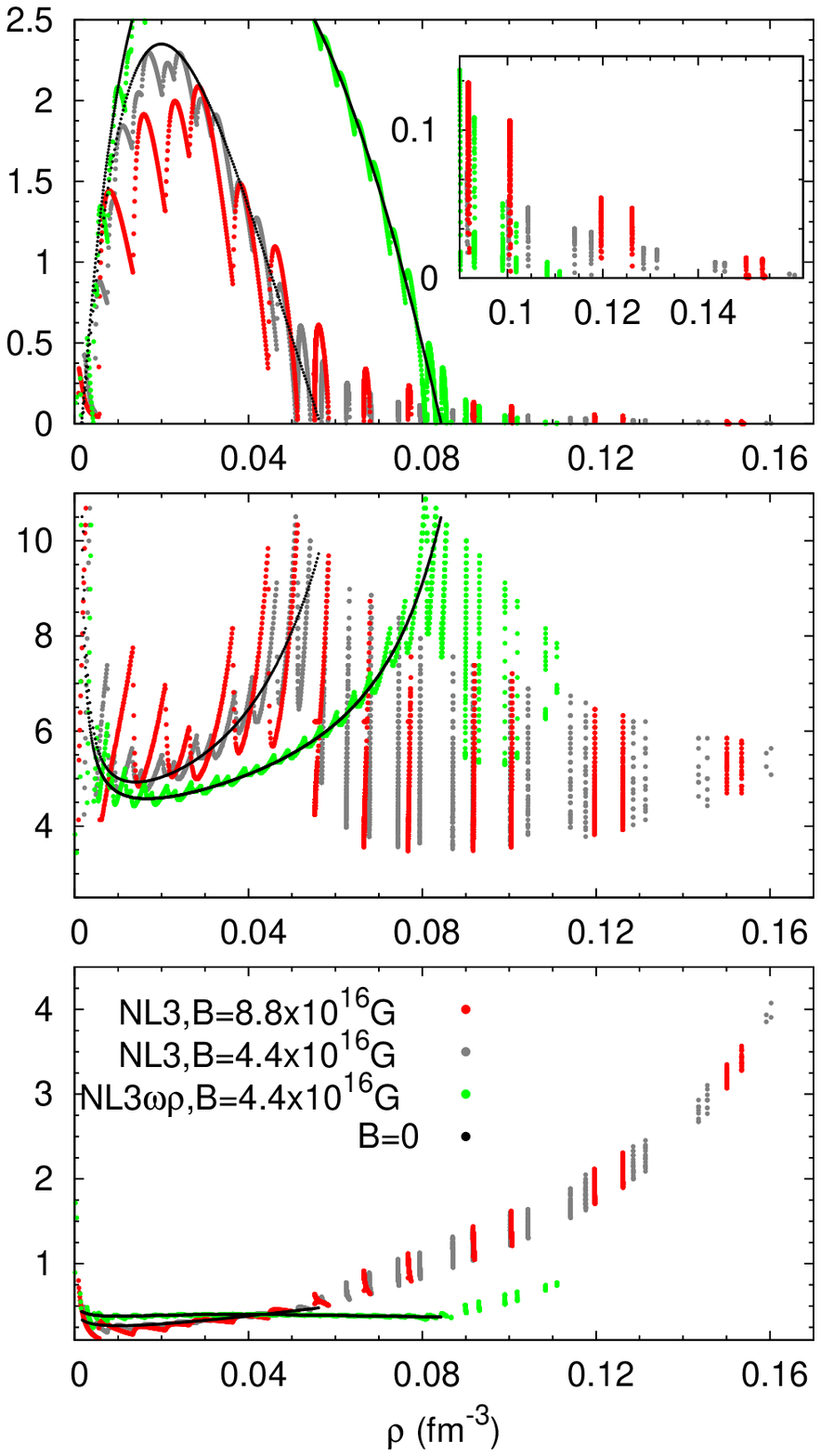} 
\end{tabular}
\caption{Largest growth rate  $\Gamma=|\omega|$ (top panels), the corresponding 
half-wavelength (middle panels) and the proton-neutron density
fluctuation ratio (bottom panels) versus density, for different magnetic field intensities and
matter with $y_p=0.02$ for NL3 and $y_p=0.035$ for NL3$\omega\rho$. The
black curve corresponds to the $B=0$ results. }
\label{rate}
\end{figure}

{ In the right panel of Fig. \ref{spin}, we can see that} a magnetic field equal to  $4.4\times10^{17}$ G 
is strong enough to
create bands of instability at densities above 0.05 fm$^{-3}$, 
associated with the filling of the different Landau levels. 
 In \cite{Rabhi2009}, the thermodynamical
  spinodal section, which corresponds to the $k=0$ limit of the dynamical
  spinodal, excluding electrons and the Coulomb field, was studied for
  magnetic fields equal or above 5$\times 10^{18}$ G. Spinodal
  bands are also present, although the stronger the field, the smaller
  the number of bands. The appearance of bands was attributed to the
  behavior of the proton chemical potential with density within each Landau band:
  at the bottom of the band it has a very soft behavior, however, at the top of the Landau level it
  hardens and a cusp occurs when a new Landau level opens, followed by
  a softening of the chemical potential.
{The proton and neutron AMM give rise to  extra bands. 
}

There are mainly two
contributions for the spinodal section: a) a closed region that contains the $B=0$  spinodal  and extra regions that form spike-like structures, associated with the filling of Landau levels;
  b) disconnected regions that appear with the opening of new Landau
  levels at densities well above the $B=0$ crust-core transition density.  
 The maximum  growth rates at constant proton fraction (see Fig. \ref{rate}) allow a more clear picture: there is a closed region that, although with some
  fluctuations, follows the $B=0$ curve, followed by separate regions,
  whose density width decreases continuously until homogeneous matter
  sets in. These disconnected regions appear when a new Landau level starts
  being filled.
 
 In \cite{Rabhi2009}, it was shown that the
  extension of the  thermodynamical spinodal for $\rho_p=0$ was
  independent of $B$, with the border to homogeneous matter at  $\rho_n^{NL3}=0.213$ fm$^{-3}$ and 
  $\rho_n^{NL3\omega\rho}=0.122$ fm$^{-3}$  for 
  the models we consider  in the present study. 
These numbers set an   upper limit of the extension of the dynamical spinodal, which in
  fact is too high, because matter in the stars has a finite proton
  fraction:
{
 for NL3 (NL3$\omega\rho$) and the proton fraction
  at the crust-core transition, $y_p^{NL3}=0.02$ ($y_p^{NL3\omega\rho}=0.035$), the dynamical
    spinodal extension is reduced to $\rho\sim 0.16$ (0.115)
    fm$^{-3}$ for $B=4.4\times 10^{16}\,G$. Decreasing further the magnetic field to 2.2
    $\times 10^{14}$G, the extension of the spinodal decreases to 0.105 fm$^{-3}$
    for NL3, and 0.102 fm$^{-3}$ for NL3$\omega\rho$,  showing a convergence to the
    $B=0$ result, respectively 0.056 and 0.084 fm$^{-3}$.}


It was shown in \cite{avancini08} that the size of the clusters in the
inner crust of a neutron star, calculated within the  TF
framework for RMF models, is well estimated by the half-wave length
associated to the most unstable mode, i.e., the one that drives matter
into a nonhomogeneous phase \cite{santos08}. In the same work, it was also shown
that for NL3 the average proton fraction in the inner crust, for densities
above 0.01 fm$^{-3}$, is $y_p\sim 0.02$, while for NL3$\omega\rho$ is $y_p=0.035$. We have, therefore, determined the most unstable modes for magnetized matter for both
parametrizations with the corresponding proton fractions, in order to
get an estimation of the size and change of charge content of the clusters
formed in the inner crust.

In Fig. \ref{rate}, the largest growth rates (top panels), the
corresponding half-wave length (middle panels) and the ratio
$\delta\rho_p/\delta\rho_n$ between the proton and neutron density
fluctuations (bottom panels) are shown for fields
between $B=2.2\times 10^{15}$ G and $8.8\times 10^{16}$ G for NL3 and NL3$\omega\rho$. In all figures,
the $B=0$ results are represented by  a black curve.
First, we consider the strongest field, represented by red dots in the right panel of
Fig. \ref{rate}, obtained only for NL3. In this case, there are several well
defined regions of clusterized matter separated by regions of
homogeneous matter.
This is a consequence
of the bands of instability, due to the filling of the Landau
levels. Also, the size of the clusters is affected. 
 In the first region of instability, the size of the
clusters oscillates around the results for $B=0$, and
fast size changes  occur in a very small density interval. 
After the first instability
region, several others appear, although the
larger the density, the smaller the density width of each region. 
Considering weaker fields, all these features are repeated with a
denser appearance of unstable regions but with smaller widths each.
The transition density to homogeneous matter is changing slowly 
with the magnetic field intensity, but considering sufficiently small
fields, the finite $B$ spinodal converges to the $B=0$ one, as
discussed before.
{This convergence is reflected on the decrease of the magnitude
    of the growth rate at a density 10\% larger than the $B=0$
    crust-core transition density with a decrease of
  the magnetic field:  for NL3 (NL3wr), it goes
    from  0.3855 (0.1481) MeV at 4.4$\times 10^{16}$ G, to  0.0921
    (0.0378) MeV  for  8.8$\times 10^{15}$ G and
  0.0068 (0.0020) MeV for 4.4$\times 10^{14}$G. }



The extension of the region with disconnected unstable
 regions is strongly dependent on the density dependence of the
 symmetry energy: for the NL3$\omega\rho$ parametrization with $L=55$
 MeV, the unstable region extends only until $\rho=0.113$
 fm$^{-3}$. This increases to $\rho\sim0.12, \, 0.13, \, 0.16$
 fm$^{-3}$ respectively for $L=68, \, 88,\, 118$ MeV and stellar
 matter conditions. Taking a
  larger proton fraction, $y_p=0.1$ which may be more realistic at
  larger densities, there will still appear unstable regions  for
  $\rho\le0.11$ fm$^{-3}$ for NL3$\omega\rho$ with $L=55$ MeV and
  $\rho\le 0.135$ fm$^{-3}$ for NL3, with $L=118$ MeV.

The proton-neutron density fluctuation ratio
was also calculated. Although $y_p=0.02$ corresponds to $\rho_p/\rho_n=1/49$, the  fluctuations
 give rise to clusterized matter with a much larger proton content:
 above the B=0 crust-core transition density the fluctuations $\delta \rho_p/\delta
  \rho_n$ increase from  $\sim0.35$ to more than the double for NL3$\omega\rho$,  and
  a factor of 5 for NL3, see both panels of Fig. \ref{rate}.

In conclusion, we have shown that  Landau
levels originate  a spinodal section with a structure of  bands,  with
disconnected regions for the
larger densities.
We expect that this
irregular border, including disconnected regions, will give rise to a more heterogeneous and amorphous phase of
matter than the one already expected due to the formation of the pasta
phases. Studies of electron conduction in this matter are needed to
confirm whether it would reduce electron conductivity, and thus originate a resistive layer within
the crust of magnetized neutron stars, as proposed in
\cite{pons13}.

The study was complemented with the determination of the largest growth rate inside the spinodal
surface, which has  allowed to estimate  the size of the clusters formed, when
the system is driven into a nonhomogeneous phase. It was shown that close to the transition to
homogeneous matter, there is a heterogeneous region, alternating
nonhomogeneous and homogeneous matter, with proton richer  clusters.
 Inside the spinodal section, the average size of the clusters  and its proton
content vary in an oscillatory way, reinforcing the heterogeneity of the inner crust
matter. 
 All these
effects are very sensitive to the density dependence of the symmetry energy.

To conclude, we refer  another pulsar property that could be affected by
an increase of the crust.
Pulsar glitches are attributed  to the angular momentum transfer
between the crust and the core \cite{link99}, involving the vortex dynamics
associated to the neutron superfluid confined to the inner crust.
However, the recent detection of an anti-glitch \cite{glitch1},  or the
indication that due to entrainment the inner crust angular moment is not
enough to explain the glitch mechanism \cite{glitch2},  suggests that the glitch
theory has to be clarified. The effects of the magnetic field on the
inner crust, in particular, an increase of the crust, the
  succession of clusterized and homogeneous layers, and a non-monotonic
  change of the neutron gas  background density will certainly affect
  the glitch mechanism and should be taken into account in a glitch theory.

\bigskip
We thank Isaac Vida\~na for useful discussions.
This work  is  partly  supported  by  the  FCT (Portugal) projects
PEst-OE/FIS/UI0405/2014, developed under the initiative QREN, and  UID/FIS/04564/2016, by
COST Action MP1304 ``NewCompStar'', and by CNPq. H.P. is supported by
FCT under Project SFRH/BPD/95566/2013.

\end{document}